\documentclass[10pt,twocolumn,aps,pra,showpacs,superscriptaddress,floatfix,nofootinbib]{revtex4-1}
\usepackage{amsmath,graphicx,epsfig,color,times}
\usepackage{pifont,bbding,amssymb, bbold}
\usepackage{indentfirst}
\usepackage{mathtools}
\usepackage{tikz,pgfplots}
\usepackage{diagbox}

\newcommand{\bra}[1]{\langle #1|}
\newcommand{\ket}[1]{|#1\rangle}

\newcommand{\GHZn}{\ket{\mathrm{GHZ}_n}}
\newcommand{\GHZ}[1]{\ket{\mathrm{GHZ}_{#1}}}
\renewcommand{\S}{\mathcal{S}}

\newcommand{\I}{\mathcal{I}}

\newcommand{\tr}{{\rm tr}}

\newcommand{\vecP}{\vec{P}}
\newcommand{\Imabk}{\I_{{\rm M}_n}}
\newcommand{\Imabkn}[1]{\I_{{\rm M}_#1}}
\newcommand{\Q}{\mathcal{Q}}

\newcommand{\IS}[1]{\I_{{\rm S}#1}}

\usepackage{comment}
\usepackage[colorlinks=true,citecolor=blue,urlcolor=blue,breaklinks]{hyperref}
\newcommand{\Npt}{N_\text{\footnotesize tr}}

\tikzset{
  treenode/.style = {shape=rectangle, rounded corners,
                     draw, align=center,
                     top color=white, bottom color=blue!20},
  root/.style     = {treenode, bottom color=red!30},
  env/.style      = {treenode},
  dummy/.style    = {circle,draw}
}
\renewcommand{\L}{\mathcal{L}}
\newcommand{\Lnmk}[3]{\L_{(#1,#2,#3)}}
\newcommand{\Qnmk}[3]{\Q_{(#1,#2,#3)}}

\usepackage[capitalize]{cleveref}
\crefformat{section}{Sec.~#2{}#1{}#3}

\begin{document}

\title{\Large Device-independent certification of multipartite entanglement using measurements performed in randomly chosen triads \\ \vspace{0.3cm}
\normalsize }
\author{Shih-Xian Yang}
\affiliation{Department of Physics and Center for Quantum Frontiers of Research \& Technology (QFort), National Cheng Kung University, Tainan 701, Taiwan}
\author{Gelo Noel Tabia}
\affiliation{Center for Quantum Technology and Department of Physics, National Tsing Hua University, Hsinchu 300, Taiwan}
\affiliation{Department of Physics and Center for Quantum Frontiers of Research \& Technology (QFort), National Cheng Kung University, Tainan 701, Taiwan}
\author{ Pei-Sheng Lin}
\affiliation{Department of Physics and Center for Quantum Frontiers of Research \& Technology (QFort), National Cheng Kung University, Tainan 701, Taiwan}
\author{Yeong-Cherng Liang}
\email{ycliang@mail.ncku.edu.tw}
\affiliation{Department of Physics and Center for Quantum Frontiers of Research \& Technology (QFort), National Cheng Kung University, Tainan 701, Taiwan}

\begin{abstract}
We consider the problem of demonstrating non-Bell-local correlations by performing local measurements in randomly chosen triads, i.e., three mutually unbiased bases, on a multipartite Greenberger-Horne-Zeilinger state. Our main interest lies in investigating the feasibility of using these correlations to certify multipartite entanglement in a device-independent setting. In contrast with previous works, our numerical results up to the eight-partite scenario suggest that if each triad is randomly but uniformly chosen according to the Haar measure, one always (except possibly for a set of measure zero) finds Bell-inequality-violating correlations. In fact, a substantial fraction of these is even sufficient to reveal, in a device-independent manner,  various higher-order entanglement. In particular, for the specific cases of three parties and four parties, our results---obtained from semidefinite programming---suggest that these randomly generated correlations always reveal, even in the presence of a non-negligible amount of white noise, the genuine multipartite entanglement possessed by these states. In other words, provided local calibration can be carried out to good precision, a device-independent certification of the genuine multipartite entanglement contained in these states can, in principle, also be carried out in an experimental situation without sharing a global reference frame.
\end{abstract}
\date{\today}
\maketitle

\section{Introduction}
    An intriguing feature of quantum theory is that, even after being separated far apart, it is still possible for distant parties sharing an appropriate entangled state to produce strongly correlated measurement outcomes~\cite{EPR}. Even more astonishingly, Bell showed that such synchronized behavior between spatially separated subsystems cannot admit a local-hidden-variable~\cite{bell}, or, more generally, a locally causal~\cite{Bell04} description---a fact that is often referred to as  (quantum) nonlocality. Importantly, such a phenomenon has now been demonstrated in a couple of so-called loophole-free Bell experiments~\cite{Hensen2015,Giustina:PRL:2015,Shalm:PRL:2015,Rosenfeld:PRL:2017}, 
    under strict locality condition in a tripartite scenario~\cite{Erven:2014aa}, as well as over a great distance~\cite{Yin:2017aa}.
    
    Following the advent of quantum information science, Bell-nonlocal~\cite{Brunner:RMP:2014} (hereafter abbreviated as nonlocal) correlations have assumed a fundamentally different role. For example, their presence signifies the security~\cite{Ekert91} of certain quantum key distribution (QKD) protocols, even when one only makes {\em minimal} assumptions~\cite{diqkd}. Similarly, Mayers and Yao~\cite{743501,MY04} found that certain extremal nonlocal correlation can be used to {\em self-test} quantum apparatus, i.e., to certify that the underlying state and the measurements employed are---modulo irrelevant degrees of freedom---essentially as expected. These findings laid the foundations of the thriving field of {\em device-independent} (DI) quantum information~\cite{Scarani12,Brunner:RMP:2014}, where nontrivial conclusions can be drawn directly from the observed data.

    It is worth noting that, although no assumption about the internal workings is needed in making a DI statement, the implementation of any protocol that relies on Bell-nonlocality still requires the spatially separated parties to perform some well-chosen local measurements. Often, this is achieved by getting the distant parties to share a reference frame---a task that is not necessarily trivial, especially if one is moving rapidly with respect to the other, as in the case of a Bell test performed between a satellite-based experimenter and a ground-based experimenter~\cite{Yin:2017aa}. 
    
    For the task of QKD, Laing {\em et al.}~\cite{PhysRevA.82.012304} have proposed a reference-frame-free protocol to circumvent the problem. In the context of demonstrating a Bell violation, a first proposal was given in Ref.~\cite{Liang:PRL:2010} to bypass this technical requirement by performing measurements in two randomly, but uniformly chosen bases. In particular, it was found that if the $n$ parties share a Greenberger-Horne-Zeilinger (GHZ) state and each chooses their two measurement bases randomly, then the chance that they would succeed in demonstrating a Bell-inequality violation increases rapidly with $n$. Moreover, this chance improves significantly~\cite{Liang:PRL:2010} if the two local measurements are further restricted to be mutually unbiased~\cite{Schwinger570,doi:10.1142/S0219749910006502}. 
    
    A couple of further investigations have since been considered. Firstly, it was shown in~\cite{Wallman:PRA:2011} that---for $n$ up to six---the findings of Ref.~\cite{Liang:PRL:2010} are robust against some local noise models. Furthermore, if the distant parties could share a {\em direction} (instead of a full reference frame) and perform their two mutually unbiased measurements on the same two-dimensional plane, then the chance of violation is provably {\em unity}. Subsequently, it was independently shown in Refs.~\cite{Shadbolt:SR:2012} and~\cite{Wallman:PRA:2012} that, even if no common direction is shared, for $n=2$, the chance of violation remains as {\em unity} if each party is allowed to perform, instead, local measurements in a {\em triad}, i.e., three mutually unbiased bases. This observation, in particular, has led to a different kind of reference-frame-free (DI) QKD protocol considered in Ref.~\cite{Slater_2014}.
    
    Besides, it was also found in Ref.~\cite{Shadbolt:SR:2012} that even without requiring the local measurements to be mutually unbiased, the probability of violation can also be boosted to (near) unity by making the number of measurement bases sufficiently large. On the other hand, Ref.~\cite{Wallman:PRA:2012} also considered the same problem for $n$ up to six, and showed numerically that not only is the probability of violation (except for the case of $n=3$) always equals unity, but the corresponding Mermin-Ardehali-Belinskii-Klyshko (MABK)~\cite{PhysRevLett.65.1838,PhysRevLett.67.2761,PhysRevA.46.5375,Belinski93}  Bell-inequality violation is also robust against white noise. More recently, Senel \emph{et al.}~\cite{FurkanSenel:PRA:2015} revisited this problem for $n=3,4,5$ and investigated (using MABK and a few other Bell-type inequalities) the probability that such randomly generated correlations would reveal either genuine $n$-partite entanglement or so-called genuine multipartite nonlocality~\cite{Svetlichny:PRA:1987}. Finally, it is worth noting that when the measurements are not restricted to be triads, some other exhaustive investigations have been carried out in the multiqubit scenario~\cite{deRosier:PRA:2017,deRosier:PRA:2020}, in the two-qudit scenario~\cite{Fonseca:PRA:2018} (see also Ref.~\cite{Barasinski2020:PhysRevA.101.052109} for an experimental demonstration in the tripartite scenario).

    Although the analysis of Ref.~\cite{FurkanSenel:PRA:2015}  is interesting, it is somewhat too restrictive because the family of MABK Bell inequalities is not the only (facet) Bell inequality defined for these Bell scenarios. In fact, even for the purpose of revealing so-called genuine $n$-partite entanglement, there is no reason to consider only facet Bell inequalities. In addition, in the event that one fails to reveal $n$-partite entanglement, it may still be possible to certify that the correlation {\em must have} originated from a quantum state with more than two-party entanglement, i.e., having an entanglement depth~\cite{ED_2001} $>2$. In this regard, we revisit the problem and extend the analysis of Refs.~\cite{Wallman:PRA:2012,FurkanSenel:PRA:2015} to (1) include the case of $n=7,8$, (2) consider the complete set of facet Bell inequalities (explicitly for the three-partite scenario, and implicitly for $n>3$), and (3) consider a general device-independent witness that is not necessarily due to a facet Bell inequality.

    In particular, we begin by explaining the concepts and the tools that we employ in~\cref{Sec:Preliminaries}. Then, our results concerning the certification of entanglement depth using specific Bell-like inequalities are presented in ~\cref{Sec:Res-ED}. Analogous results obtained without resorting to particular Bell inequalities together with their white-noise robustness are summarized in~\cref{Sec:Res-Vis}. We conclude with a discussion and possible future directions in ~\cref{Sec:Conclusion} while leaving miscellaneous details to the appendixes.

\section{Preliminaries} \label{Sec:Preliminaries}
\subsection{Various sets of correlations and their membership test}
\label{Sec:Membership}

We now introduce concepts that are relevant to the current investigation. Consider an $n$-partite Bell experiment where each party has a choice over $m$ measurement settings and where each measurement results in one of $k$ possible outcomes. 
We denote the {\em correlation}, i.e., the conditional probability distributions of observing outcomes $\vec{a}_n:= (a_1,a_2,\dots,a_i,\dots,a_n)$ given settings $\vec{x}_n:= (x_1,x_2,\dots,x_i,\dots,x_n)$ by $\vec{P}:= \{P(\vec{a}_n|\vec{x}_n)\}_{\vec{a}_n,\vec{x}_n}$; here $x_i$ and $a_i$ are, respectively, the label of the measurement setting chosen and the measurement outcome observed by the $i$-th party. Throughout, we use the notation $(n,m,k)$ to refer to the Bell scenario being considered. For instance, $(3,3,2)$ refers to a Bell scenario involving three parties, and with each of them performing three dichotomic measurements. In this work, we shall only focus on Bell scenarios $(n,2,2)$ and $(n,3,2)$, where $n=3,4,\ldots,8$. For concreteness, the labels are then assumed to take the values of $a_{i}\in\{0,1\}$, $i\in\{1,2,\ldots,n\}$, $x_{i}\in\{0,1\}$, and $x_i\in\{0,1,2\}$, respectively for $m=2$ and $m=3$.

Depending on the resource shared by the parties, the set of correlations $\vec{P}$ that they can generate would have to satisfy different mathematical constraints. For example, if the parties only have access to shared randomness, then $\vec{P}$ satisfies:
\begin{equation}
  \label{eq:LocalSet}
  P(\vec{a}_n|\vec{x}_n) = \sum_{\lambda}P(\lambda)\prod_{i=1}^n \delta_{a_i,f_i(x_i,\lambda)}, 
\end{equation}
for some choice of local response functions $f_i(x_i,\lambda)$ and some weight $P(\lambda)\ge 0$ such that $\sum_\lambda P(\lambda)=1$. Correlations satisfying~\cref{eq:LocalSet} are said to be Bell-local~\cite{Brunner:RMP:2014} (hereafter abbreviated as local) and we denote the set of all these $\vecP$ as $\L$ (or $\Lnmk{n}{m}{k}$ if we want to be precise about the exact symmetric Bell scenario involved). It is worth noting that, for finite $n,m,k$, there are only a finite number of deterministic strategies $f_i(x_i,\lambda)$ --- $\L$ is thus a (convex) polytope~\cite{Brunner:RMP:2014}. 

On the other hand, if a quantum state $\rho$ is shared by the participants and the correlation is generated by them performing a local measurement on their respective subsystem, then according to Born's rule, $\vec{P}$ takes the form of
\begin{equation}
  \label{eq:Q}
  P(\vec{a}_n|\vec{x}_n) = \tr\left( \rho \bigotimes_{i=1}^n M_{a_i|x_i}\right), 
\end{equation}
where $\{M_{a_i|x_i}\}_{a_i,x_i}$ is the positive-operator-valued-measure representing the $x_i$-measurement of the $i$th party. We denote the set of such conditional quantum distributions by $\Q$ (or by $\Qnmk{n}{m}{k}$ if we want to be precise about the Bell scenario involved). It is well known that $\L\subset\Q$ and as was first shown by Bell~\cite{bell}, the inclusion is strict.

To make it evident that a given $\vecP\not\in\L$, one can employ a \emph{witness}, called a Bell inequality~\cite{bell}, which can be written without loss of generality as
\begin{equation}
  \label{eq:BellIneq}
\I(\vec{P}) :=  \sum_{\vec{a},\vec{x}}\beta_{\vec{a}}^{\vec{x}}P(\vec{a}|\vec{x}) \overset{\L}{\leq} I_{\mathcal{L}} \overset{\Q}{\leq} I_{\mathcal{Q}}. 
\end{equation}
This means that for all $\vec{P} \in \L$, the value of the linear combination of $P(\vec{a}_n|\vec{x}_n)$  specified by $\beta_{\vec{a}}^{\vec{x}}$ is upper bounded by $I_\L$. As a result, if one observes a value of $\I(\vecP)$ greater than $I_\L$, it must be that $\vec{P}\not\in\L$, and this conclusion follows regardless of how $\vec{P}$ is generated from the underlying state and measurements. This independence from the internal workings of the device is the basis of so-called device-independent (DI) quantum information~\cite{Scarani12}, where one draws nontrivial conclusions about the nature of the employed devices directly from the observed data.

In a similar manner, one can also consider more refined separations arising from the differences in the many-body entanglement possessed by the shared quantum resource~\cite{Curchod:PRA:2015}. For instance, one may require that the shared state $\rho$ is $k$-producible~\cite{Guhne_2005}, i.e., $\rho$ can be written as a convex combinations of $k$-producible pure states: 
\begin{equation}  \label{kproducible}
    \ket{\psi} = \ket{\phi^{(1)}} \otimes \ket{\phi^{(2)}} \otimes \dots \otimes \ket{\phi^{(m)}}, 
\end{equation}
where each tensor factor $\ket{\phi^{(i)}}$ involves \emph{at most} $k$ parties. From this definition, it follows that a $k'$-producible state is also $k$-producible for all $k\ge k'$. Hence if we denote by $\Q_{n,k}$ the set of correlations obtainable via Eq.~\eqref{eq:Q} when $\rho$ is $n$-partite but $k$-producible ($1\le k\le n$),\footnote{The notation for a quantum $k$-producible set, which contains only two subscripts and without any brackets, is not to be confused with that for the full quantum set $\Qnmk{n}{m}{k}$ defined for the  specific Bell scenario $(n,m,k)$.} then $\L=\Q_{n,1}\subseteq \Q_{n,2}\subseteq \cdots \subseteq\Q_{n,n-1}\subseteq \Q$. Here, the first equality follows from the fact that a one-producible state is fully separable and such states cannot~\cite{Werner:PRA:1989} violate any Bell inequality. On the other hand, the last inclusion follows from the fact that, in an $n$-partite scenario, the set of $n$-producible quantum states is simply the set of all $n$-partite quantum states.

As a result, if we denote by $I_{k\text{-prod.}}$ the maximal value of $\I(\vecP)$ [cf. Eq.~\eqref{eq:BellIneq}] attainable by $\vecP\in\Q_{n,k}$, then 
\begin{equation}\label{eq:kproduciblebound}
  \I(\vec{P}) \overset{\L}{\leq} I_{\text{1-prod.}} \overset{\Q_{n,2}}{\leq} I_{\text{2-prod.}}  \le \cdots \overset{\Q_{n,n-1}}{\leq} I_{\text{($n$-1)-prod.}}\overset{\Q}{\leq} I_{\Q}. 
\end{equation}
Thus, in analogy to the idea of witnessing a nonlocal correlation $\vec{P}$ using a Bell inequality, if a value greater than $I_{\text{($k$-1)-prod.}}$ is observed,  the underlying quantum state $\rho$ \emph{cannot} be $(k-1)$-producible. In particular, a quantum state that is $k$-producible but {\em not} $(k-1)$-producible is said to have an {\em entanglement depth} of $k$~\cite{ED_2001}. Consequently, an inequality like
\begin{equation}\label{Eq:DIWED}
    \I(\vecP)\overset{\Q_{n,k}}{\le} I_{\text{$k$-prod.}}
\end{equation} 
is said to be a device-independent witness for entanglement depth (DIWED)~\cite{Liang:PRL:2015} because it allows one to certify that the shared state must have an entanglement depth (ED) of $k+1$ or more (see, e.g., Refs.~\cite{Nagata:PRL:2002,Yu:PRL:2003,Bancal:PRL:2011,Liang:PRL:2015,Lin:PRA:2019,Aloy:PRL:2019,Tura:PRA:2019} for some explicit examples).

Crucially, since the labels of the measurement settings $x_i$, the measurement outcomes $a_i$, and even the party $i$ are arbitrary, one can start from any given DIWED [cf. Eq.~\eqref{Eq:DIWED}] and generate a different, but {\em equivalent} DIWED via relabeling. For example, one may apply to $\beta^{\vec{x}}_{\vec{a}}$ the permutation of label $i=1\leftrightarrow i=n$, as well as $a_i=0\leftrightarrow a_i=1$ to some (or all) of the measurement outcomes. With some thought, it should be clear that the resulting inequality is still a valid DIWED for all $\vecP\in\Q_{n,k}$. As such, one may start from Eq.~\eqref{Eq:DIWED} and generate an entire family of other {\em equivalent} DIWEDs $\I_j(\vecP)=\sum_{\vec{a},\vec{x}}\Pi_j(\beta^{\vec{x}}_{\vec{a}}) P(\vec{a}|\vec{x})\overset{\Q_{n,k}}{\le} I_{\text{$k$-prod.}}$ by simply applying a permutation $\Pi_j$ on these labels attached to $\beta^{\vec{x}}_{\vec{a}}$. 
Since each of these DIWEDs is satisfied by all $\vecP\in\Q_{n,k}$, the set of $\vecP$ satisfying such a family of DIWEDs would form a {\em polytopic} superset of $\Q_{n,k}$.

Indeed, for any given quantum correlation $\vecP$, the violation of a given DIWED (or any equivalent DIWED obtained from relabeling) is not the only means to lower-bound the underlying entanglement depth. In particular, as explained in Appendix G of Ref.~\cite{Liang:PRL:2015}, deciding if a given $\vecP$ lies in $\Q_{n,k}$ can be achieved by solving a hierarchy of semidefinite programs (SDPs), each giving a tighter {\em outer} approximation of $\Q_{n,k}$. Let us denote the  $\ell$-th level outer approximation of $\Q_{n,k}$  (see Refs.~\cite{Liang:PRL:2015,Moroder:PRL:2013}) by $\S^{(\ell)}_{n,k}$, i.e., $\Q_{n,k}\subseteq \S^{(\infty)}_{n,k}\subseteq\cdots \subseteq\S^{(\ell)}_{n,k}\subseteq\cdots\subseteq \S^{(2)}_{n,k}\subseteq \S^{(1)}_{n,k}$, then it is worth noting that each $\S^{(\ell)}_{n,k}$ is {\em convex} but generally {\em not} polytopic. For any given $\vecP$, its membership with respect to $\S^{(\ell)}_{n,k}$ (and hence to  $\Q_{n,k}$) can be decided by solving the following SDP:
    \begin{equation}
    \label{eq:vis}
    \begin{split}
     \sup &\qquad\qquad\qquad v \\
    {\rm s.t.}&\;\; \vecP(v):= v \vec{P} + (1-v)\vecP_w\;\in\; \S^{(\ell)}_{n,k};  \\
    &\qquad\qquad\qquad v \ge 0,
    \end{split}
    \end{equation}
where $\vec{P}_w$ is the {\em white noise}, i.e., the {\em uniform} probability distribution, and the membership test with respect to $\S^{(\ell)}_{n,k}$ requires only the implementation of matrix positivity constraints. 

Note that $v=0$ is always a feasible solution to the SDP since $\vec{P}_w\in\L=\Q_{n,1}\subseteq\S^{(\ell)}_{n,k}$ for all $\ell$ and all $k\ge 1$. From the convexity of $\S^{(\ell)}_{n,k}$, it follows that if $\vecP\in\S^{(\ell)}_{n,k}$, the optimum value to the problem, denoted by $v^*$, satisfies $v^*\ge 1$.
On the other hand, the convexity of $\S^{(\ell)}_{n,k}$ also implies that $v^*$ must be strictly less than one whenever $\vecP\not\in\S^{(\ell)}_{n,k}$. In other words, if $v^*$ (often referred to as the {\em white-noise visibility}, or simply visibility) is less than one for any $\ell$, then $\vecP\not\in\S^{(\ell)}_{n,k}\supset \Q_{n,k}$, and thus the quantum state giving rise to $\vecP$ must have an ED of $k+1$ or higher. This is the tool that allows us to go beyond the investigation of Refs.~\cite{Wallman:PRA:2012,FurkanSenel:PRA:2015}, which considers only specific Bell inequalities or DIWEDs.

    \begin{figure}[h!]
    \includegraphics{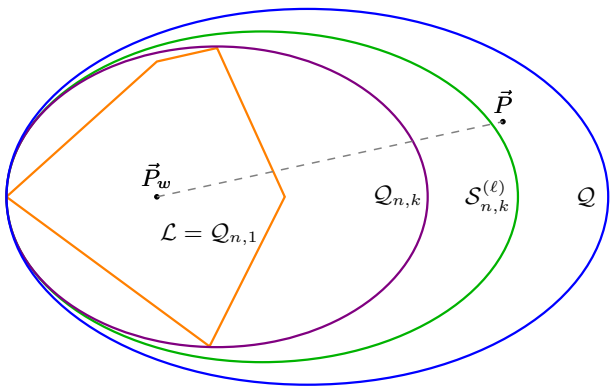}
	\caption{\label{fig:producible_set}
	Schematic illustration of a $k$-producibility test achieved via the correlation $\vecP$ obtained in a Bell test. The test relies on the observation that if $\vecP\not\in\S^{(\ell)}_{n,k}\supseteq\Q_{n,k}$, then the locally measured quantum state $\rho$ is not $k$-producible. The figure represents a two-dimensional projection of the space of all possible $\vecP$. Membership of $\vecP$ with respect to the local set $\L=\Q_{n,1}$ (orange polygon) can be decided by either solving a linear program or when $\vecP$ is found to violate a (facet) Bell-inequality (orange edge). Analogously,  $\vecP$ is known to lie outside $\Q_{n,k}$ (violet oval) for $k>1$ if it is found to violate a DIWED $\Q_{n,k}$ or if the optimum value $v^*$ obtained by solving Eq.~\eqref{eq:vis} with respect to some relaxation $\S^{(\ell)}_{n,k}$ (green oval) of $\Q_{n,k}$ is found to be less than one. 
s}
   	\end{figure}
   	
    For comparison, let us also note that whenever a DIWED such as Eq.~\eqref{Eq:DIWED} is violated by some $\vecP'$, a corresponding white-noise visibility with respect to this witness can be computed [cf. Eq.~\eqref{eq:vis}] as:
    \begin{equation}\label{Eq:VisIneq}
        v = \frac{\I(\vecP_w)-I_{\text{$k$-prod.}}}{\I(\vecP_w)-\I(\vecP')}.  
    \end{equation}
    If the local measurements leading to this violation gives, instead, $\vecP_w$ when acting on the maximally mixed state $\frac{\mathbb{1}}{2^n}$, this visibility can also be understood~\cite{Grandjean:PRA:2012} as the infimum of the $v$ needed for the state
    	\begin{equation}\label{Eq:NoisyGHZ}
    	    \rho(v) = v\GHZn\bra{\mathrm{GHZ}_n} + (1-v)\frac{\mathbb{1}}{2^n}
    	\end{equation}
    to violate the given witness for the \emph{very same} local measurements. Evidently, the visibility $v$ obtained from Eq.~\eqref{Eq:VisIneq} is always larger than or equal to the visibility obtained by solving Eq.~\eqref{eq:vis} because the latter involves an implicit optimization over all possible witnesses. These concepts are illustrated schematically in~\cref{fig:producible_set}.   	
Before concluding this section, let us also recall that, for relatively simple Bell scenarios, the membership test of $\L=\Q_{n,1}$ can be carried out exactly (rather than relying on a membership test of outer approximations) as the problem reduces to a linear program over a convex polytope. 

\subsection{Probability of certifying entanglement depth $\geq k$}
\label{Sec:ProbViolation}

To investigate the feasibility of certifying entanglement, and, more generally, the correct entanglement depth by performing measurements in randomly chosen triads, we need to investigate if the resulting correlations are \emph{always outside} the relevant $k$-producible sets. To this end, we follow the procedure of Ref.~\cite{Shadbolt:SR:2012} but consider its extension to more than two parties. Suppose we have $n$ parties that share a
GHZ state $\GHZ{n} =\frac{1}{\sqrt{2}}\left(\ket{0}^{\otimes n} + \ket{1}^{\otimes n}\right)$. Each party can perform a set of three mutually unbiased qubit measurements. Because such a set corresponds to three orthogonal directions on the Bloch sphere, we call it a \emph{triad}. Here we focus on correlations obtained from triads that are chosen independently and uniformly at random.

Since every mutually unbiased observable associated with a triad can be obtained by performing a unitary transformation on the Pauli observables $\sigma_x$, $\sigma_y$, and $\sigma_z$, we may without loss of generality, sample a qubit unitary instead of directly sampling a triad on the Bloch sphere. Hence, to sample a triad uniformly at random, each party picks a Haar-random unitary matrix and applies it to measurements of the three Pauli observables. A Haar-random unitary is generated by sampling a matrix from the complex Ginibre ensemble \cite{doi:10.1063/1.1704292}, performing a QR  decomposition on that matrix, and multiplying each column of $Q$ by the sign of the corresponding diagonal entry of $R$ \cite{Mezzadri2007}.  In this case, every choice of $n$ independent random unitaries produces a single correlation $\vecP\in\Qnmk{n}{3}{2}$.

Let an $n$-triad set be a set of $n$ triads associated with a quantum correlation $\vecP$. From $\vecP$, we want to determine whether $\vecP$ could have arisen from an underlying quantum state that is $k$-producible.
Roughly, if we define a uniform distribution over all possible $n$-triad sets, or equivalently all possible choices of $n$ independently sampled qubit unitaries, then the probability $p^{(n,k)}$ of certifying ED $\ge k$ would be given by the fraction of $n$-triad sets whose corresponding $\vecP$ are {\em certified} to be outside $\Q_{n,k-1}$. It should be pointed out that $p^{(n,k)}$ is only a {\em lower bound} on the probability of finding a randomly sampled $\vecP$ that lies outside $\Q_{n,k-1}$. This is because our certification, as explained in ~\cref{Sec:Membership}, makes use of outer approximations of $\Q_{n,k-1}$, either via $\S^{(\ell)}_{n,k-1}$ or via the polytopic superset of $\Q_{n,k-1}$ obtained from specific DIWEDs (more on this below).
Formally,
\begin{equation}
    p^{(n,k)} = \int f(\Omega)\,{\rm d}\Omega
\end{equation}
where ${\rm d}\Omega$ represents the Haar measure over $n$ independently chosen  qubit unitaries and $f(\Omega)$ is an indicator function that returns one if the unitaries corresponding to $\Omega$ yields a correlation $\vecP$ that is certified to lie outside $\Q_{n,k-1}$ but vanishes otherwise. 

As explained in~\cref{Sec:Membership} (see, e.g., Figure~\ref{fig:producible_set}), there are two different ways to certify that a given $\vecP$ lies outside $\Q_{n,k-1}$: either by solving Eq.~\eqref{eq:vis} and finding $v^*<1$, or evaluating a DIWED $\I(\vecP)$ [cf. Eq.~\eqref{Eq:DIWED}] and finding that it is violated. Obviously, when the latter approach is invoked, it can only help to consider not just a single DIWED, but also all of its equivalent forms obtained from an arbitrary relabeling. Therefore, whenever we invoke a specific DIWED, i.e., a Bell-like inequality (equipped with the relevant $k$-producible bound $I_\text{$k$-prod.}$) to perform such a certification (as in ~\cref{Sec:Res-ED}), it goes without saying that all its equivalent forms obtained from relabeling are also considered at the same time. In other words, we do not test any individual DIWED, but rather the polytopic superset of $\Q_{n,k-1}$ that results from the DIWED of interest.

To obtain an estimate of $p^{(n,k)}$, we therefore perform repeated trials for $\Npt$ times and compute the relative frequency of trials whereby the corresponding $\vecP$ is certified to be outside $\Q_{n,k-1}$. Additionally, we can approximate the probability density function by plotting a histogram of the corresponding visibilities, using appropriately chosen bin widths.

\subsection{Three paths for certifying the (non) $k$-producibility of $\vecP$}
\label{Sec:3Paths}

As mentioned above, since we consider local measurements on a triad, our sampled correlation $\vecP$ is defined for the Bell scenario $(n,3,2)$. It is thus most natural to perform the relevant membership test in this Bell scenario. However, since very little is known in relation to Bell inequalities (let alone DIWEDs) with three measurement settings~\cite{Laskowski2004:PRL.93.200401,PhysRevLett.95.120405,Zukowski2006}, when we perform a membership test by considering specific Bell inequalities or DIWEDs, we shall consider exclusively only Bell inequalities that are naturally defined for the $(n,2,2)$ scenario. Clearly, we can still test our sampled correlation $\vecP$ against all these inequalities defined for a Bell scenario with one less measurement setting: by disregarding all entries of $\vecP$ pertaining to one of the measurement settings of each party, one obtains $\vecP_\text{sub}\in\Q_{n,2,2}$.

For completeness, we should nonetheless consider all $\left[{3}\choose{2}\right]^n=3^n$ ways of selecting two measurements from each triad and determine the combination that gives the largest Bell value, and hence the optimal visibility, i.e., the smallest value of $v$ according to Eq.~\eqref{Eq:VisIneq}. In doing so, we effectively consider all possible \emph{input-liftings}~\cite{Pironio_Lifting} of a Bell inequality or a DIWED---initially defined for the $(n,2,2)$ Bell scenario---to the $(n,3,2)$ Bell scenario. Again, we emphasize that, when lifting a Bell inequality or DIWED, we implicitly take into account all its equivalent forms obtained from relabeling.

In a similar manner, each of these different $\vecP_\text{sub}$ may be subjected to a membership test in the $(n,2,2)$ Bell scenario by using Eq.~\eqref{eq:vis}. Exploiting the terminology of {\em lifting} introduced for Bell inequalities, we shall refer to the best visibility obtained in this manner as the visibility with respect to the lifting of $\S^{(\ell)}_{n,k}$ to the $(n,3,2)$ Bell scenario (see also~\cite{Chellasamy:PRR:2019}). Importantly, liftings generally give rise to only a subset of all legitimate Bell inequalities (or DIWEDs) defined for the $(n,3,2)$ Bell scenario. The optimum visibility obtained in this manner is therefore generally suboptimal compared with that determined directly by solving Eq.~\eqref{eq:vis} with $\vecP\in\Q_{n,3,2}$.

\begin{figure}
        \centering
        \includegraphics{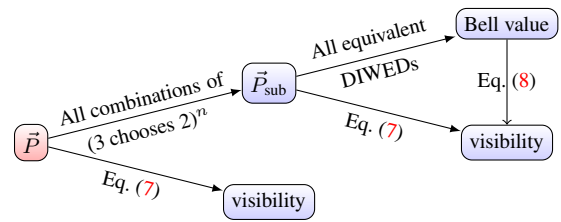}
        \caption{\label{fig:scenarios}Flowchart summarizing the three different approaches employed in this work for certifying the (non) $k$-producibility of a sampled $\vecP\in\Q_{n,3,2}$. In the uppermost branch, we first obtain $\vecP_\text{sub}\in\Q_{n,2,2}$ by keeping only the relevant entries of $\vecP$. All $3^n$ such $\vecP_\text{sub}$ obtained from a single $\vecP$ are then tested against a DIWED and all its equivalent forms to determine the largest Bell value, and hence the optimal visibility via~\cref{Eq:VisIneq}. Effectively, this uppermost branch tests $\vecP$ against all {\em} input liftings of DIWEDs originally defined for the $(n,2,2)$ scenario. Secondly, for each $\vecP_\text{sub}$, we  solve~\cref{eq:vis} to determine the corresponding optimal visibility $v^*$; the smallest of all these $3^n$ visibilities then gives the visibility of $\vecP$ with respect to the lifting of $\S^{(\ell)}_{n,k}$ in the $(n,2,2)$ scenario to the $(n,3,2)$ Bell scenario. This is graphically represented as a sub-branch of the upper branch. Finally, in the bottom branch, we solve~\cref{eq:vis} directly for the optimal visibility $v^*$ with respect to $\S^{(\ell)}_{n,k}$ defined for the $(n,3,2)$ Bell scenario. 
}

    \end{figure}

These three paths for determining the $k$-producibility of $\vecP$ are summarized in \cref{fig:scenarios}.

\section{DI certification using specific Bell inequalities}
\label{Sec:Res-ED}

We now assess the behavior of our randomly sampled correlations by evaluating particular Bell inequalities. In this section, we focus  on the lifting of $\S^{(\ell)}_{n,k}$ in the $(n,2,2)$ scenario to the $(n,3,2)$ scenario, i.e., picking two measurements from each randomly chosen triad and keeping the combination of $n$ pairs that yields the largest Bell value among a family of equivalent DIWEDs.

\subsection{The MABK Bell Inequality}
 First we consider the $n$-partite MABK inequality $\Imabk$~\cite{PhysRevLett.65.1838,PhysRevLett.67.2761,PhysRevA.46.5375,Belinski93}, where $\GHZ{n}$ is known to exhibit a maximal Bell violation that is exponential in $n$.\footnote{Note that a strengthened version of the MABK inequality that achieves also an exponential violation can be found in Ref.~\cite{Kai:PRA:2006}.} 
 Here we wish to highlight our results for the cases $n=7,8$, which expands upon the analysis of previous works~\cite{Wallman:PRA:2012,FurkanSenel:PRA:2015}. 
The MABK Bell inequality can be written in the compact form~\cite{Wallman:PRA:2011}:
    \begin{equation}\label{eq:mabk}
        \Imabk(\vec{P}) = {\sum_{\vec{x}\in \{0,1\}^n} \beta(\vec{x}_n)E(\vec{x}_n)}\overset{\L}{\leq} 1,         
    \end{equation}
    where $E(\vec{x}_n)=\sum_{\vec{a}_n\in \{0, 1\}^n} \prod_{i=1}^n (-1)^{a_i} P(\vec{a}_n|\vec{x}_n)$, and the coefficients $\beta(\vec{x}_n)$ are given by
    \begin{equation}
    \beta(\vec{x}_n) = 2^{\frac{1-n}{2}} \cos\left[ \frac{\pi}{4}(1+n-2x)\right],
	\end{equation}
    where $x=\sum_{i=1}^{n} x_i$. 

Numerically, we find that the probability of witnessing nonlocal correlations by using the family of $n$-partite MABK inequalities is \emph{unity} in all cases except the tripartite case. This is consistent with the observation made in Ref.~\cite{Wallman:PRA:2012} (but not with Ref.~\cite{FurkanSenel:PRA:2015} for the $n=4$ case) and extends it to the scenarios with $n=7,8$. In addition, we note that the probability of violating the two-producible bound of $\Imabk$ is also unity for $n>4$.

\begin{figure}
        \centering
\includegraphics{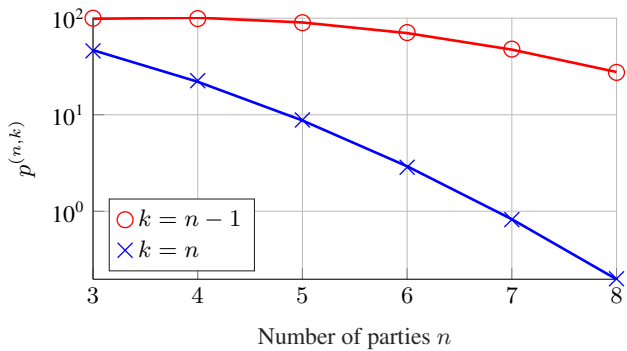}
    \caption{\label{fig:mabk-n_ent} Semilog plot (base 10) of the probability of certifying ED $n$ (blue) and $n-1$ (red) as a function of the number of parties $n$. The curves are our best (quadratic) fit of the numerical data. Explicitly, the red curve admits the expression $\log_{10}p^{(n,n-1)} = -0.0368n^2-0.0695n+2.2089$ whereas that of the blue curve is $\log_{10}p^{(n,n)} = -0.0304n^2+0.2240n+1.5955$.}
    \end{figure}

However, in contrast with the claim in~\cite{FurkanSenel:PRA:2015}, we observe that the chance of witnessing the GME nature of $\GHZn$ decays rapidly with the number of parties $n$.\footnote{In particular, for $n=5$, we observe a 8.83\% probability of violating the corresponding DIWED whereas Ref.~\cite{FurkanSenel:PRA:2015} reported 19\% for the corresponding probability.}
In a similar fashion, the chance of certifying an ED $\ge n-1$ is also seen to decrease exponentially with increasing $n$. 
This suggests that, while $\Imabk$ is useful in detecting the entanglement of $\GHZn$, it is rather ineffective in revealing the exact entanglement depth of these states in the present context.
We summarize our results for the probability of violating $k$-producible bounds of $\Imabk$ in~\cref{tbl:ProbkPB_MABK} and provide the fitting function of certifying ED to be $(n-1)$ and $n$, respectively, in~\cref{fig:mabk-n_ent} .

    \begin{table}[hbt!]
    \centering
    \begin{tabular}{|c|c|c|c|c|c|c|c|} 
    \hline
    $\Npt~ (10^6)$ &  4 & 4 & 2 & 0.467 & 0.276 &
    0.125 \tabularnewline \hline \hline 
    \diagbox{$k$}{$n$} & 3 & 4 & 5 & 6 & 7 & 8 \tabularnewline \hline
    1  & $^*$100  & 100  & 100  & 100  & 100  & 100  \tabularnewline \hline
    2 & 45.89  & 99.10  & 100  & 100  & 100  & 100   \tabularnewline \hline
    3&-& 22.54  & 89.84  &NA& 99.28  & 99.99  	\tabularnewline \hline
    4&-&-&8.83 & 70.98  & NA&NA	\tabularnewline \hline
    5&-&-&-& 2.86 & 47.84 & NA	\tabularnewline \hline
    6&-&-&-&-& 0.82  & 27.45    	\tabularnewline \hline
    7&-&-&-&-&-&0.20   \tabularnewline \hline
    \end{tabular}
    \caption{\label{tbl:ProbkPB_MABK} Summary of the probability of violating the $k$-producible bounds (and hence witnessing an ED of at least $k+1$) using $\Imabk$ for $n=3,4,\ldots,8$. 
    The first row describes the number of random correlations sampled for each scenario,  $\Npt$. Note that entries marked as $^*100$ are those where some instances of {\em no} violation have been found but they represent less than $0.01\%$ of the total samples. Entries marked with ``NA'' are cases where the $k$- and $(k+1)$-producible bounds overlap (see Table~\ref{tbl:kPB}).}
    \label{mabk}
    \end{table}

\subsection{Other facet Bell inequalities in $(3,2,2)$}

    Given that the MABK Bell inequality alone is insufficient to always reveal the entanglement (depth) of $\GHZ{3}$ in the current setting, it is natural to ask if there exist other tripartite Bell inequalities that are more suited for this task. To this end, it is worth noting that, in the $(3,2,2)$ Bell scenario, the complete set of facet Bell inequalities characterizing the local set $\Lnmk{3}{2}{2}$ has been determined by Sliwa in Ref.~\cite{Sliwa:PLA:2003,Sliwa:arXiv}. After taking into account the freedom in relabeling, these facet inequalities can be classified into 46 inequivalent families (with $\IS{2}$ being equivalent to the 3-partite MABK inequality $\Imabkn{3}$), but only 44 of these can be violated in quantum theory. From the results of Ref.~\cite{Vallins:PRA:2017}, it follows that only 25 families of inequalities display a gap of more than 10$^{-5}$ between their maximal quantum violation and their two-producible bounds.
    
   By testing our $\Npt=10^6$ randomly sampled correlations against the 44 potentially useful Bell inequalities, we identify 11 for which the local bound is apparently {\em always} violated. These are $\IS{4}$, $\IS{5}$, $\IS{6}$, $\IS{8}$, $\IS{16}$, $\IS{22}$,  $\IS{24}$, $\IS{28}$, $\IS{33}$, $\IS{39}$, and $\IS{42}$. On the other hand, aside from the positivity facet $\IS{1}$  and the guess-your-neighbor-input (GYNI) inequality $\IS{10}$~\cite{PhysRevLett.104.230404}, we find three other facet Bell inequalities (namely, $\IS{3}$, $\IS{11}$, and $\IS{23}$) that {\em never} seem to be violated. Among those 41 inequalities that can be used to reveal the entanglement of the $\GHZ{3}$, 11 of them can even be used---in a probabilistic manner---to reveal its tripartite entanglement by performing local measurements on these randomly chosen triads. In particular, the 33rd inequality in the list of Sliwa exhibits a significant advantage over the MABK inequality in terms of certifying the correct entanglement depth of $\GHZ{3}$ from these correlations (see Table~\ref{Table:SliwaFull}). Even then, none of the DIWEDs arising from these Bell facets is, by itself, always sufficient to certify the GME nature from these randomly sampled correlations. In fact, even if we take the intersection defined by all of them---which forms again a polytopic relaxation of $\S_{3,2}$---the probability of success in this task can only be boosted to approximately 68.97\%, which is about 7\% better compared with considering $\I_{S33}$ alone.
    \begin{table}[ht]
    \begin{tabular}{|c|r|r|r|r|r|r|r|r|r|r|r|r|r|r|r|r|r|r|r|r|} 
    \hline
     $k$ & $\I_{\rm S1}$ & $\I_{\rm S2}$& $\I_{\rm S3}$ & $\I_{\rm S4}$ & $\I_{\rm S5}$ & $\I_{\rm S6}$ & $\I_{\rm S7}$ & $\I_{\rm S8}$ & $\I_{\rm S9}$ & $\I_{\rm S10}$ 
     \\\hline\hline
      1 & 0 & $^*$100& 0 & 100 & 100 & 100 & 99.85 & 100 & $^*100$ & 0
     \\ \hline
     2 & 0 & 45.89 & 0 & 0 & 0 & 0 & 6.82 & 4.97 & 0 & 0 \\  \hline\hline
     
     $k$ & $\I_{\rm S11}$ & $\I_{\rm S12}$& $\I_{\rm S13}$ & $\I_{\rm S14}$ & $\I_{\rm S15}$ & $\I_{\rm S16}$ & $\I_{\rm S17}$ & $\I_{\rm S18}$ & $\I_{\rm S19}$ & $\I_{\rm S20}$ \\ \hline \hline 
     1 & 0 & 22.75 & 99.98 & 97.06 & 98.33 & 100 & 99.97 & 40.91 & 99.92 & 96.62 \\ \hline
     2 & 0 & 0 & 0 & 0 & 0 & 0 & 0 & 0 & 0 & 0\\ \hline\hline    
     
     $k$ & $\I_{\rm S21}$ & $\I_{\rm S22}$& $\I_{\rm S23}$ & $\I_{\rm S24}$ & $\I_{\rm S25}$ & $\I_{\rm S26}$ & $\I_{\rm S27}$ & $\I_{\rm S28}$ & $\I_{\rm S29}$ & $\I_{\rm S30}$     \\\hline\hline
     1 & $^*100$ & 100 & 0 & 100 & 99.39 & 96.36 & $^*100$ & 100 & 99.96 & $^*100$ 
     \\ \hline
     2 & 0 & 0.18 & 0 & 16.63 & 0 & 3.98 & 0.67  & 0 & 0 & 0\\ \hline\hline    
     
     $k$ & $\I_{\rm S31}$ & $\I_{\rm S32}$& $\I_{\rm S33}$ & $\I_{\rm S34}$ & $\I_{\rm S35}$ & $\I_{\rm S36}$ & $\I_{\rm S37}$ & $\I_{\rm S38}$ & $\I_{\rm S39}$ & $\I_{\rm S40}$     \\\hline\hline
     1 & 66.06 & 99.89 & 100 & 99.32 & 86.02 & 99.96 & 99.91 & 98.36 & 100 & 99.76 \\ \hline
     2 & 0 & 0 & 61.92 & 0 & 0 & 0 & 0  & 0 & 39.20 & 2.31\\ \hline\hline        
     
     $k$ & $\I_{\rm S41}$ & $\I_{\rm S42}$& $\I_{\rm S43}$ & $\I_{\rm S44}$ & $\I_{\rm S45}$ & $\I_{\rm S46}$ &  &  &  &    \\\hline\hline
     1 & $^*100$ & 100 & 99.58 & 99.83 & 99.95  & 91.15 &  &  &  & 
     \\ \hline
     2 & 0 & 3.08 & 0 & 0 & 0 & 0 &  &  &  &  \\ \hline\hline   
     \end{tabular}     %
    \caption{\label{Table:SliwaFull} Summary of the probability of violating the $k$-producible bounds ($k=1,2$) of all 46 facet Bell inequalities in the $(3,2,2)$ Bell scenario~\cite{Sliwa:PLA:2003,Sliwa:arXiv}. The $i$th inequality in this list is denoted by $\IS{i}$. In particular, $\IS{1}$ and $\IS{10}$ represent, respectively, the positivity facet and the GYNI inequality~\cite{PhysRevLett.104.230404},
    both of which are known to be satisfied by quantum theory. These probabilities are expressed in percent and were estimated by using a total of $\Npt=10^6$ samples, except for $\I_{\rm S 2}$, which was estimated by using $\Npt=4\times10^6$ samples.  The first column gives the value of $k$ while the entries for each inequality are summarized in a single column, spanning across two rows for the two different values of $k$. Entries marked as $^*100$ are those where some instances of {\em no} violation have been found but they represent less than $0.01\%$ of the total samples.}
    \end{table}
    
\subsection{Some other Bell inequalities in $(n,2,2)$} \label{other_Bell}
\label{subsec:otherBelln22}

For $n>3$ parties, little is known in terms of the complete set of facets. However, since our goal is to investigate the effectiveness of using nonlocal correlations to reveal the entanglement depth of $n$, it would make sense to investigate how some other Bell inequalities---known to be violated by $\GHZn$---fare in the current task.

The first candidate in our list is a natural generalization of Sliwa's seventh inequality $\IS{7}$, first introduced in Ref.~\cite{Liang:PRL:2015} for an {\em arbitrary} number of parties:
	\begin{equation}
	\label{Ineq:Sliwa7}
	\I_{{\rm S}_n}(\vec{P})=2^{1-n} \left(\sum_{\vec{x}\in \{0,1\}^n} E_n(\vec{x}_n)\right)  - E_n(\vec{1}_n) \overset{\L}{\leq} 1. 
	\end{equation} 
For $n\le 8$, this has been established to be a facet inequality of the local set $\Lnmk{n}{2}{2}$~\cite{Liang:PRL:2015}. Additionally, it  was shown therein that the $k$-producible bounds of $\I_{{\rm S}_n}$  coincide with the maximal quantum violation of $\I_{{\rm S}_k}$, as long as $k\le n$. The numerical values of these $k$-producible bounds for $k\le 8$ are known~\cite{Liang:PRL:2015}, and are partially reproduced in Table~\ref{tbl:kpbofS7FG} for ease of reference. Notice that there is a nontrivial gap between the $(k-1)$- and $k$-producible bounds of $\I_{{\rm S}_n}$  for any $k \leq 8$ (we only reproduce the bounds for $k\le 6$ in the table).

Our second candidate is again a family of Bell inequalities that are known to be maximally violated by $\GHZn$, or equivalently~\cite{Hein_2004} (under the freedom of local unitaries) the fully-connected graph states. Explicitly, these inequalities proposed in Ref.~\cite{Lin:PRA:2019} read as
	\begin{align}
	\nonumber
	\I_{{\rm FG}_n}(\vec{P}) &= \tfrac{1}{n-1} \left[ E_n(0,1,1, \ldots, 1) + \circlearrowright^{'} - E_n(\vec{0}_3, \vec{1}_{n-3}) \right] \\ & \overset{\L}{\leq} 1. 
	\label{Ineq:FG}
	\end{align} 
    The first term indicates that all parties, except the first one, perform the zeroth measurement. The symbol, $\circlearrowright^{'}$,  stands for the additional $n-1$ terms that have to be included so that the first $n$ terms becomes invariant under a cyclic permutation of parties. 
    Note that $\I_{{\rm FG}_3}$ is equivalent to $\I_{\mathrm{M}_{3}}$.
    
The probability of correctly certifying the ED of $\GHZn$ using the aforementioned Bell inequalities is summarized in~\cref{tbl:ResultS7FG}. For fixed $n$, we find that the chance of detecting the nonlocality with randomly generated correlations is higher when using $\I_{{\rm S}_n}$. The fact that $\I_{{\rm S}_n}$ is a facet of $\Lnmk{n}{2}{2}$ while $\I_{{\rm FG}_n}$ is not could have played a role in this difference. However, the numerical results also indicate that, for certifying ED of $k>2$, $\I_{{\rm FG}_n}$ is clearly preferable to $\I_{{\rm S}_n}$. For instance, when $n=5$, the probability of witnessing ED $k\ge 3$ vanishes for $\I_{{\rm S}_5}$, but for $\I_{{\rm FG}_5}$ it is low but nonetheless nonzero. In any case, we see that when it comes to certifying the correct ED of $\GHZn$ (for $n\ge 4$), both these families of inequalities are far inferior when compared with DIWEDs arising from $\Imabk$ (see Tables~\ref{tbl:ProbkPB_MABK} and \ref{tbl:ResultS7FG} for details).

    \begin{table}[h!bt]
	\centering
	\begin{tabular}{|c |c |c |c |c |c |}
    \hline
    $\Npt~(10^6)$ & 1 & 0.28 & 0.01 & 1 & 0.029 \tabularnewline
    \hline \hline
    \diagbox{$k$}{$\I$} & $\I_{{\rm S}_4}$& $\I_{{\rm S}_5}$ & $\I_{{\rm S}_6}$ &  $\I_{{\rm FG}_4}$ &   $\I_{{\rm FG}_5}$ \tabularnewline    \hline 
    1  & 99.08 & 75.50 & 26.24 & 91.13 & 54.01 \tabularnewline \hline 
    2  & 0.39 & 0 & 0 & 32.48 & 12.57 \tabularnewline\hline
    3  & 0 & 0  & 0 & 1.3 & 0.82 \tabularnewline \hline
    4  & - & 0 & 0 & - & 0.17  \tabularnewline \hline
    5  & - & - & 0 & - & - \tabularnewline\hline
    \end{tabular}
	\caption{\label{tbl:ResultS7FG} Summary of the probability of violating the $k$-producible bounds ($1\le k\le n-1$) of inequality $\I_{\rm S_n}$, \cref{Ineq:Sliwa7} for $n=4,5,6$ and the inequality $\I_{{\rm FG}_n}$ tailored for the fully-connected graph state, \cref{Ineq:FG} for $n=4,5$. These probabilities are expressed in percent and the number of sampled correlations, $\Npt$, are listed in the first row.}
	\end{table}

\section{DI certification using membership tests and White-noise Robustness}
\label{Sec:Res-Vis}

\begin{figure}
    \centering
    \includegraphics{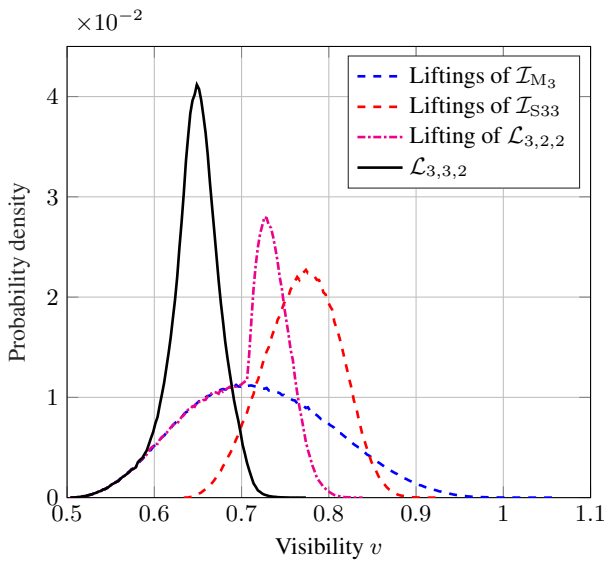}
    \caption{Visibility distribution of the randomly sampled correlation $\vecP$ to the local polytopes in the $n=3$ case. Included here are the visibility distribution obtained from the Bell value optimized over all input liftings of $\Imabkn{3}$  (or equivalently, $\IS{2}$, dashed blue line), all input liftings of $\IS{33}$ (dashed red line), that obtained by solving Eq.~\eqref{eq:vis} with $k=1$ when we only consider the best combination of two out of the three measurement settings (i.e., the input-lifting of $\Lnmk{3}{2}{2}$, dashed-dotted magenta line), and when we consider all the three measurement settings together ($\Lnmk{3}{3}{2}$, solid black line). We have $\Npt=10^6$ for these histograms except the data for $\Imabkn{3}$, in which we have $\Npt = 4\times 10^6$.  All histograms are plotted with a bin width of 2.5$\times 10^{-3}$.}
	\label{fig:VisToLocal:n3}
\end{figure}

\begin{figure}
	    \centering
	    \includegraphics{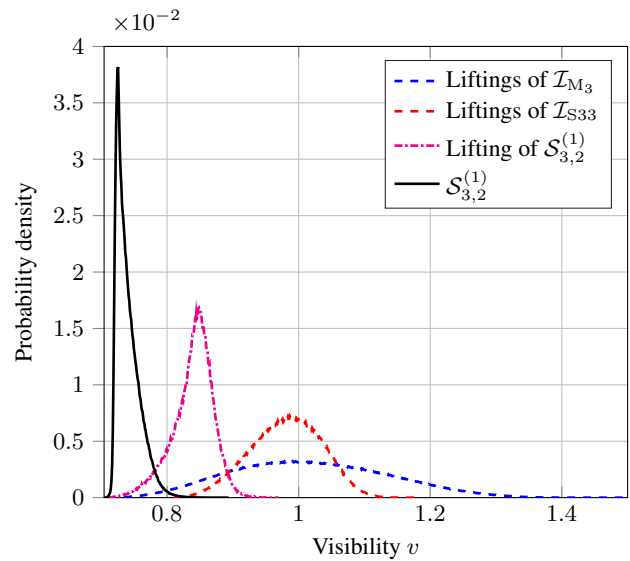}
	    \caption{\label{fig:vis-to-2prod} Visibility distribution of the randomly sampled correlation $\vecP$ to the two-producible set $\Q_{3,2}$, via four different approximations in the $n=3$ case. 
Included here are the visibility distribution obtained from the Bell value optimized over all input liftings of $\Imabkn{3}$  (or equivalently, $\IS{2}$, dashed blue line), all input liftings of $\IS{33}$ (dashed red line), that obtained by solving~\cref{eq:vis} with $k=2$ when we only consider the best combination of two out of the three measurement settings (lifting of $\S_{3,2}^{(1)}$, dashed-dotted magenta line), and when we consider all the three measurement settings together ($\S_{3,2}^{(1)}$, solid black line). $\Npt=2.5\times 10^5$ for the lifting of $\S_{3,2}^{(1)}$,  $\Npt = 10^6$ for $\S_{3,2}^{(1)}$ and the liftings of $\IS{33}$, whereas $\Npt=4\times 10^6$ for the liftings of $\Imabkn{3}$. All histograms are plotted with a bin width of 0.001.}
	\end{figure}

In the previous section, we discuss the behavior of our randomly sampled correlations in terms of a few specific families of Bell inequalities. 
Here we give the corresponding probabilities obtained by 
performing membership tests on the sampled $n$-partite correlations with respect to the local set $\L$ and the $(n-1)$-producible set $\Q_{n,n-1}$. As explained in ~\cref{Sec:Preliminaries}, this is achieved by solving Eq.~\eqref{eq:vis} by using the sampled $\vecP$. In particular, for the membership test with respect to $\L$, Eq.~\eqref{eq:vis} reduces to a linear program whereas for the membership test with respect to $\Q_{n,n-1}$, we make use the first-level outer approximation $\S_{n,n-1}^{(1)}$ of $\Q_{n,n-1}$ in our computation. Apart from being able to check against all Bell inequalities (in the $k=1$ case) and all DIWEDs (in the $k=n-1$ case) at the same time for the appropriate Bell scenario, such membership tests also immediately give the white-noise robustness of these correlations. Our results for these membership tests for the $n=3$ case with $k=1$ and $k=2$ are illustrated, respectively, in Figs.~\ref{fig:VisToLocal:n3} and \ref{fig:vis-to-2prod}. For comparison, the visibility distributions obtained by evaluating the Bell value of some specific Bell inequalities discussed in~\cref{Sec:Res-ED}, in accordance to~\cref{Eq:VisIneq}, are also displayed in these figures.

Interestingly, even though some of the Bell facets of $\Lnmk{3}{2}{2}$, such as $\IS{33}$, fares better in terms of the probability of Bell violation, they are {\em not} generally better than $\Imabkn{3}$ in terms of white-noise robustness. Indeed, it is clear from Figure~\ref{fig:VisToLocal:n3} that if we admix $\GHZ{3}$ with a sufficiently larger amount of white noise (e.g., if $v\le 0.62$), then $\IS{33}$ can no longer be violated but $\Imabkn{3}$ can still be violated with some nonzero probability. The overlapping curves might even suggest that when testing the randomly sampled correlations against {\em all} lifted Bell facets  of $\Lnmk{3}{2}{2}$ together, all those contributing to smaller visibilities, say, with $v<0.71$, are due {\em entirely} to a violation of $\Imabkn{3}$. Moreover, this general feature still holds  (albeit with a smaller critical visibility of $v\approx 0.58$) even if we consider all Bell facets of $\Lnmk{3}{3}{2}$ together. From Table~\ref{tbl:lbvis_state} we see that these results are also fairly robust against depolarizing noise: if we consider only the Bell facets lifted from $\Lnmk{3}{2}{2}$, the probability of a Bell violation would stay as unity even if up to 18\% of  white noise is present; if we consider, instead, all (including those nonlifted) Bell facets of $\Lnmk{3}{3}{2}$, then this white noise tolerance can be improved to about 22\%. 

What about the certification of the GME nature of $\GHZ{3}$ using these randomly sampled correlations? We mentioned in the last section that even by considering {\em all} the DIWEDs derived from the Bell facets of $\Lnmk{3}{2}{2}$ and lifting them to the $(3,3,2)$ Bell scenario, the probability of witnessing an entanglement depth of three is still far from unity. However, via the membership test of Eq.~\eqref{eq:vis}, we see from Fig.~\ref{fig:vis-to-2prod} that it is apparently always possible to certify the GME nature of $\GHZ{3}$ by using these randomly sampled correlations. In fact, as can be seen from Table~\ref{Tbl:Vis-2-sep}, such a certification is robust, as the probability of success remains as unity even if we allow the presence of about 3.1\% of white noise and inspect only two out of the three local measurement settings at one time. If all three measurement settings are considered together, then this white-noise tolerance can be boosted to about 10.7\% (see Table~\ref{Tbl:Vis-2-sep}).

\begin{figure}
    \centering
    \includegraphics{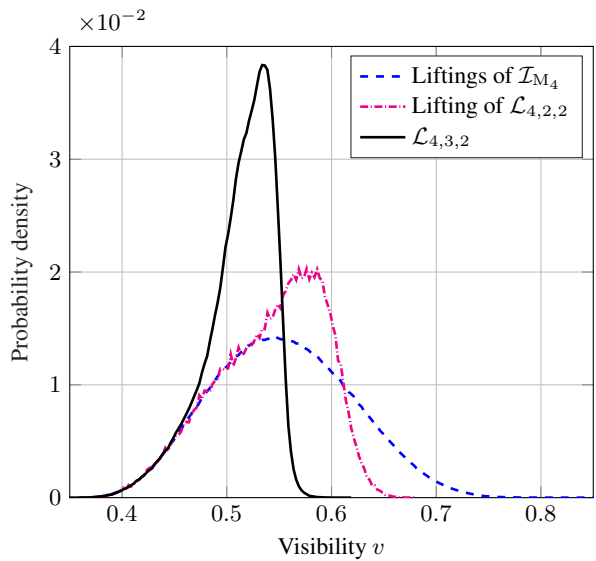}
	\caption{\label{fig:VisToLocal:n4}Visibility distribution of the randomly sampled correlation $\vecP$ to the local polytopes in the $n=4$ case. Included here are the visibility distribution obtained from the Bell value optimized over all input liftings of $\Imabkn{4}$ (dashed blue line) and that obtained by solving Eq.~\eqref{eq:vis} with $k=1$ when we only consider the best combination of two out of the three measurement settings (i.e., the input-lifting of $\Lnmk{4}{2}{2}$, dashed-dotted magenta line), and when we consider all the three measurement settings together ($\Lnmk{4}{3}{2}$, solid black line). $\Npt=4\times 10^6$ for the liftings of $\Imabkn{4}$, $\Npt=10^5$ for the lifting of $\Lnmk{4}{2}{2}$, and $\Npt=8\times 10^5$ for $\Lnmk{4}{3}{2}$. All histograms are plotted with a bin width of 0.01.}
\end{figure}

Continuing to the $n=4$ case, we see from Fig.~\ref{fig:VisToLocal:n4} that even if we restrict ourselves to considering only $\Imabkn{4}$, our ability to certify the nonlocality of $\GHZ{4}$ is already fairly robust against depolarizing noise---the probability of violation remains unity even if we admix $\GHZ{4}$ with about 12\% of white noise (see Table~\ref{vis_MABK}). Not surprisingly, this noise-resistance can be boosted considerably, leading to approximately 32\% and 38.0\% if we consider, respectively, all Bell facets of $\Lnmk{4}{2}{2}$ lifted to the Bell scenario of $(4,3,2)$ and the consideration of all Bell facets of $\Lnmk{4}{3}{2}$ (see Table~\ref{tbl:lbvis_state} for a summary of these visibility distributions).

\begin{figure}
    \centering
    \includegraphics{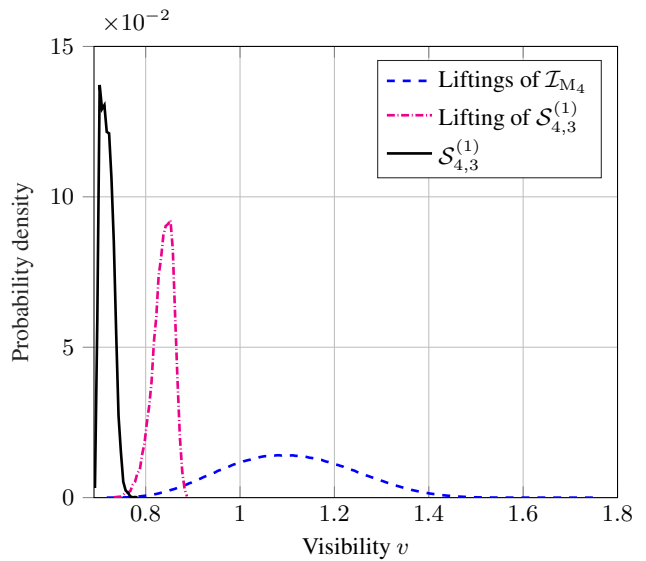}
\caption{\label{fig:Vis-to-3-prod}
Visibility distribution of the randomly sampled correlation $\vecP$ to the three-producible set $\Q_{4,3}$, via three different approximations in the $n=4$ case. Included here are the visibility distribution obtained from the Bell value optimized over all input liftings of $\Imabkn{4}$ (dashed blue line), that obtained by solving~\cref{eq:vis} with $k=3$ when we only consider the best combination of two out of the three measurement settings (lifting of $\S_{4,3}^{(1)}$, dashed-dotted magenta line), and when we consider all the three measurement settings together ($\S_{4,3}^{(1)}$, black solid line). $\Npt=1.66\times 10^4$ for the input-lifting of $\S_{4,3}^{(1)}$, $\Npt = 10^4$ for $\S_{4,3}^{(1)}$, and $\Npt = 10^4$ for the liftings of $\Imabkn{4}$. All histograms are plotted with a bin width of 0.005.}
\end{figure}

On the other hand, results from the last section (see, e.g., Fig.~\ref{fig:mabk-n_ent} and Table~\ref{tbl:ResultS7FG}) may suggest that it is unlikely to perform a DI certification of the correct ED of $\GHZ{4}$ using correlations obtained by measuring $\GHZ{4}$ in randomly sampled  triads. However, if we base our certification on solving Eq.~\eqref{eq:vis}, it is clear from the visibility distributions shown in Figure~\ref{fig:Vis-to-3-prod} that not only can we certify the correct ED of $\GHZ{4}$ with certainty, the same can also be said with the mixed state of Eq.~\eqref{Eq:NoisyGHZ} with at least 11\% of white noise. In fact, if we make use of the correlations for all the measurement settings together, then this white-noise robustness can even reach 21\%. Details in relation to these levels of white-noise tolerance can be found in Table~\ref{Tbl:Vis-2-sep}.

For $n\ge 5$ parties, it is clear from Table~\ref{vis_MABK} and ~\cref{tbl:lbvis_state} that, for the demonstration of nonlocality, or equivalently, for the DI certification of entanglement of $\GHZn$, the  protocol becomes increasingly robust, at least, for $n$ up to 8. On the other hand, for the DI certification of the GME nature of $\GHZn$, as mentioned above, a consideration based only on specific Bell inequalities is inconclusive. Unfortunately, an investigation of Eq.~\eqref{eq:vis} for a statistically significant number of samples is computationally too expensive to be carried out.

\section{Conclusion}
\label{Sec:Conclusion}

    In this work, we have---building upon the analysis of Refs.~\cite{Shadbolt:SR:2012,Wallman:PRA:2012,FurkanSenel:PRA:2015}---investigated the feasibility of demonstrating Bell-nonlocal correlations by having $n$ parties performing their measurements in a randomly chosen triad (i.e., three mutually unbiased bases) on a shared $\GHZn$. Our results for the $n=7,8$ scenarios are consistent with the trend already observed in Ref.~\cite{Wallman:PRA:2012} for $n=2,\ldots,6$. Namely, not only that such a device-independent entanglement certification protocol is (in principle) feasible, but is even strongly robust to the presence of white noise. Furthermore, when appropriate Bell inequalities are considered, we could also get around the insufficiency of the MABK Bell inequality discovered in Refs.~\cite{Wallman:PRA:2012,FurkanSenel:PRA:2015} for the $n=3$ case. In fact, even for $10^7$ randomly chosen sets of triads, we have always found the resulting correlations to violate 11 of the facet Bell inequalities defined in this tripartite, two-setting, two-outcome Bell scenario. 

    Given these encouraging observations, a natural question that arises is whether these randomly generated nonlocal correlations would be strong enough to also reveal the (genuine) multipartite entanglement contained in $\GHZn$. To this end, we have not only repeated the analysis of Ref.~\cite{FurkanSenel:PRA:2015} for the cases of $n=3,4,$ and 5 but have also analyzed the cases for $n=6,7,$ and 8 based on device-independent witnesses for entanglement depth (DIWED) obtained from the MABK Bell inequalities. In this regard, we remark that although our results for the $n=3$ case appear to agree, the findings of Ref.~\cite{FurkanSenel:PRA:2015} do not seem to be consistent with ours nor that of Ref.~\cite{Wallman:PRA:2012} for the $n=4$ case (in terms of the probability of certifying the nonlocality of $\GHZ{4}$), neither do the results of Ref.~\cite{FurkanSenel:PRA:2015} agree with ours in terms of the probability of correctly certifying the entanglement depth of $\GHZ{5}$. Unfortunately, since the raw data of Ref.~\cite{FurkanSenel:PRA:2015} is no longer available~\cite{markham2019}, we are not able to precisely pinpoint the source of this discrepancy.

    In any case, for the DI certification of entanglement depth, our results show that, if we are to consider only DIWEDs that are based on MABK Bell inequalities, then the probability of correctly certifying the entanglement depth of $\GHZn$ appears to decrease exponentially with $n$. In fact, the same conclusion holds even if we only wish to certify that its entanglement depth is larger than or equal to $n-1$. Also, for the $n=3$ case, even if we are to consider all DIWEDs constructed from the Bell facets defined for the $(3,2,2)$ Bell scenario (see Refs.~\cite{Sliwa:arXiv,Vallins:PRA:2017}), the probability of correctly certifying the entanglement depth of $\GHZ{3}$ is still less than 70\%. However, if we are willing to consider also all possible DIWEDs (including those that stem from non-facet-defining Bell inequalities)---something that we achieved by solving appropriate semidefinite programs first discussed in~\cite{Liang:PRL:2015}---then not only can we certify the correct entanglement depth with certainty, but such a certification is even robust to the presence of white noise. To our astonishment, this robustness even increases when the number of parties is increased from $n=3$ to $n=4$. As such, we conjecture that for an arbitrary number of parties, the entanglement depth of $\GHZn$ can always be certified in a DI manner using the protocol that we have discussed here.
    
    A few other remarks are now in order. First, we mentioned in~\cref{Sec:Res-ED} that, somewhat surprisingly, among all the randomly generated correlations, none of them have violated the 3rd, 11th, and the 23rd inequality presented by Sliwa~\cite{Sliwa:arXiv}. However, it is important to note that this observation is, more a feature of the nature of the measurements chosen rather than that of the state itself. In fact, if we do not impose the measurements to be mutually unbiased, one can easily find a quantum violation of all these Bell inequalities by $\GHZ{3}$. 
    
    Second, in the work of T\'oth {\em et al.}~\cite{Toth:PRA:2005}, it was pointed out that if the two local measurements involved are {\em assumed} to be orthogonal (on the Bloch sphere), then a violation of the $\Imabkn{3}$ Bell inequality itself is already sufficient to certify genuine three-qubit entanglement. Let us, nonetheless, remark that, in our analysis, although we make use of mutually unbiased measurements to generate random correlations for our analysis, in deciding whether a certification of the correct entanglement depth is successful, we have never relied on this assumption regarding the nature of the measurements, as that would render the conclusion device-{\em dependent}, rather than being device-{\em independent}.
    
    Let us now comment on some possibilities for future research. First, the current analysis, as with many other closely related work (see, e.g., Refs.~\cite{Liang:PRL:2010,Wallman:PRA:2011,FurkanSenel:PRA:2015,deRosier:PRA:2017,Fonseca:PRA:2018,deRosier:PRA:2020}) suffer from the drawback that the results presented are mostly numerical. Consequently, our observations are only known to be applicable to relatively simple Bell scenarios. To this end, it would be desirable to obtain analytic results that could, e.g., reveal the asymptotic behavior involving a large number of subsystems etc.
    Also, while our semidefinite programming approach has enabled us to correctly certify the entanglement depth of $\GHZ{3}$ and $\GHZ{4}$, it requires full knowledge of the generated correlation $\vecP$, rather than only its values with respect to certain DIWEDs. In a real experimental setting, even if we disregard various imperfections, due to statistical fluctuations, the observation of a $\vecP\in\Q$ is in practice never available (see Ref.~\cite{Lin:PRA:2018} for a discussion). For a realistic feasibility analysis of this device-independent certification protocol, statistical fluctuations must thus be taken into account, e.g., by the tools discussed in Ref.~\cite{Liang:Entropy:2019}. This is, however, clearly outside the scope of the present work and will be left to future research.

\section{Acknowledgement}
    We thank Jebarathinam Chellasamy for useful discussions.
    This work is supported by the Foundation for the Advancement of Outstanding Scholarship and the Ministry of Science and Technology, Taiwan (Grants No. 104-2112-M-006-021-MY3, No. 107-2112-M-006-005-MY2, and 107-2627-E-006-001, and 108-2627-E-006-001).

\bibliography{DIMECRandMeas}

\appendix

 \section{$k$-producible bounds of various Bell expressions}
 
    For ease of reference, we provide below the $k$-producible bounds of the Bell expressions that have been invoked in~\cref{Sec:Res-ED}. To begin, we recall from Ref.~\cite{Liang:PRL:2015} the $k$-producible bounds for the MABK Bell expression of Eq.~\eqref{eq:mabk} in Table~\ref{tbl:kPB}. 

    \begin{table}[h!bt]
    \centering
    \begin{tabular}{|c|c|c|c|c|c|c|c|} 
        \hline
        \diagbox{$k$}{$n$} &3 & 4 & 5 & 6 & 7 & 8 \tabularnewline \hline
        3&2& 2&2 &$2\sqrt{2}$& $2\sqrt{2}$ & $2\sqrt{2}$ 	\tabularnewline \hline
        4&-&$2\sqrt{2}$&$2\sqrt{2}$& $2\sqrt{2}$& 4& $4\sqrt{2}$	\tabularnewline \hline
        5&-&-&4& 4& 4&$4\sqrt{2}$	\tabularnewline \hline
        6&-&-&-&$4\sqrt{2}$& $4\sqrt{2}$& $4\sqrt{2}$ 	\tabularnewline \hline
        7&-&-&-&-&8&8  \tabularnewline \hline
        8&-&-&-&-&-&$8\sqrt{2}$  \tabularnewline \hline
    \end{tabular}
        \caption{\label{tbl:kPB} Summary of the quantum $k$-producible bounds (for $2<k \le n$) of the MABK Bell expression of Eq.~\eqref{eq:mabk} for $n=3,4,\ldots,8$. In all these cases, the omitted local bound is always 1 whereas the $2$-producible bound is always $\sqrt{2}$.}
        \end{table}
    
    Next, we reproduce the $k$-producible bounds of the Bell expressions due to Sliwa~\cite{Sliwa:PLA:2003, Sliwa:arXiv} in Table~\ref{Table:SliwaFull:kPB}. For definiteness, these bounds are applicable to the negative of the expression given in the right-hand side of the Table 1 in Ref.~\cite{Sliwa:arXiv} and ignoring the constant term. Observe that the one-producible bound, i.e., the local bound, is given by the constant term in Table 1 of Ref.~\cite{Sliwa:arXiv}. The corresponding two-producible bound (which coincides with the biseparable bound in the tripartite case) is extracted from the largest entry among the second-last to the fourth-last column of Table 1 of~\cite{Vallins:PRA:2017}.
    
    \begin{table}[h!t]
    \begin{tabular}{|c|r|r|r|r|r|r|r|r|r|r|r|r|r|r|r|r|} 
    \hline
     $k$ & $\I_{\rm S1}$ & $\I_{\rm S2}$& $\I_{\rm S3}$ & $\I_{\rm S4}$ & $\I_{\rm S5}$ & $\I_{\rm S6}$ 
     \\\hline\hline
        1 & 1 & 2 & 2 & 2 & 3 & 3 
     \\ \hline
     2 & 1 & 2$\sqrt{2}$ & 2$\sqrt{2}$ & $4\sqrt{2}-2$ & $4\sqrt{2}-1$ & $4\sqrt{2}-1$ \\  \hline\hline
     
     $k$ & $\I_{\rm S7}$ & $\I_{\rm S8}$ & $\I_{\rm S9}$ & $\I_{\rm S10}$ & $\I_{\rm S11}$ & $\I_{\rm S12}$
     \\\hline\hline
     1 & 4 & 4 & 4 & 4 & 4 & 4 \\ \hline
     2 & $4\sqrt{2}$ & $4\sqrt{2}$ & $4\sqrt{2}$ & 4 & $4\sqrt{2}$ & $4\sqrt{2}$ \\ \hline\hline
     
     $k$ & $\I_{\rm S13}$ & $\I_{\rm S14}$ & $\I_{\rm S15}$ & $\I_{\rm S16}$ & $\I_{\rm S17}$ & $\I_{\rm S18}$ \\ \hline \hline 
     1 & 4 & 4 & 4 & 4 & 4 & 4  \\ \hline
     2 & $4\sqrt{2}$ & $4\sqrt{2}$ & $4\sqrt{2}$ & $4\sqrt{2}$ & $4\sqrt{2}$ & $4\sqrt{2}$ \\ \hline\hline  
     
     $k$ & $\I_{\rm S19}$ & $\I_{\rm S20}$ & $\I_{\rm S21}$ & $\I_{\rm S22}$& $\I_{\rm S23}$ & $\I_{\rm S24}$ \\ \hline\hline
     1 & 4 & 4 & 4 & 4 & 4 & 5 \\ \hline
     2 & $4\sqrt{2}$ & $6\sqrt{2}-2$ & $4\sqrt{2}$ & $4\sqrt{2}$ & $\frac{3}{2}(\sqrt{17}-1)$ & $4\sqrt{2}+1$ \\ \hline\hline
     
     $k$ & $\I_{\rm S25}$ & $\I_{\rm S26}$ & $\I_{\rm S27}$ & $\I_{\rm S28}$ & $\I_{\rm S29}$ & $\I_{\rm S30}$     \\\hline\hline
     1 & 5 & 5 & 5 & 6 & 6 & 6 
     \\ \hline
     2 & $4\sqrt{2}+1$ & $\frac{29+\sqrt{832}}{9}$  & $4\sqrt{2}+1$  & $8\sqrt{2}-2$ & $8\sqrt{2}-2$ & $8\sqrt{2}-2$\\ \hline\hline    
     
     $k$ & $\I_{\rm S31}$ & $\I_{\rm S32}$& $\I_{\rm S33}$ & $\I_{\rm S34}$ & $\I_{\rm S35}$ & $\I_{\rm S36}$      \\\hline\hline
     1 & 6 & 6 & 6 & 6 & 6 & 6 \\ \hline
     2 & $4\sqrt{2}+2$ & $4\sqrt{2}+2$ & $4\sqrt{2}+2$ & $4\sqrt{2}+2$ & $\frac{38+\sqrt{832}}{9}$ & $8\sqrt{2}-2$  \\ \hline\hline  
     
     $k$ & $\I_{\rm S37}$ & $\I_{\rm S38}$ & $\I_{\rm S39}$ & $\I_{\rm S40}$ & $\I_{\rm S41}$ & $\I_{\rm S42}$ \\ \hline\hline
     1 & 6 & 6 & 6 & 6 & 7 & 8 \\ \hline
     2 & $8\sqrt{2}-2$ & $8\sqrt{2}-2$ & $4\sqrt{2}+2$ & $4\sqrt{2}+2$ & $8\sqrt{2}-1$ & $8\sqrt{2}$ \\ \hline\hline
     
     $k$ & $\I_{\rm S43}$ & $\I_{\rm S44}$ & $\I_{\rm S45}$ & $\I_{\rm S46}$ &  &     \\\hline\hline
     1 & 8 & 8 & 8 & 10 &  & 
     \\ \hline
     2 & $8\sqrt{2}$ & $12\sqrt{2}-4$ & $12\sqrt{2}-4$ & $12.9852$ &  & \\ \hline   
     \end{tabular}     %
    \caption{\label{Table:SliwaFull:kPB} Summary of the quantum $k$-producible bounds of the various Bell expressions due to Sliwa~\cite{Sliwa:arXiv}. Note that $\IS{7}=4\I_{{\rm S}_3}$, whereas $\IS{2}=2\Imabkn{3}$ upon relabeling the measurement settings. All analytic expressions presented are approximations of the numerical bounds given in Table 1 of Ref.~\cite{Vallins:PRA:2017}, with an accuracy that is at least $10^{-6}$. Note that the maximal quantum violation of $\IS{4}$ given in Table 1 of Ref.~\cite{Vallins:PRA:2017} was off by a factor of two, i.e., it should correspond to the two-producible bound of $4\sqrt{2}-2$ that we list here.}
    \end{table}
    
Finally, we recall in Table~\ref{tbl:kpbofS7FG} the $k$-producible bounds of $\I_{{\rm S}_n}$ from Ref.~\cite{Liang:PRL:2015} and that of $\I_{{\rm FG}_n}$ from Ref.~\cite{Lin:PRA:2019}. Notice that the $k$-producible bounds of $\I_{{\rm S}_n}$ {\em only} depend on $k$ but not on $n$.
\begin{table}[h!tbp]
	\begin{tabular}{|c|c|c|c|c|}
    	\hline
    	\diagbox{$k$}{$\I$} & $\I_{{\rm S}_n}$ & $\I_{{\rm FG}_4}$ & $\I_{{\rm FG}_5}$ & $\I_{{\rm FG}_6}$ \tabularnewline \hline 
    	2 & $\sqrt{2}$ & 1.2247 & 1.1547 & 1.1180 \tabularnewline \hline 
    	3 & $\frac{5}{3}$ & 1.4679 & 1.2291 & 1.2195 \tabularnewline \hline
    	4 & 1.8428 & $\frac{5}{3}$ & 1.3509 & 1.2392 \tabularnewline \hline
    	5 & 1.9746 & - & 1.5 & 1.2807 \tabularnewline \hline
    	6 & 2.0777 & - & - & 1.4 \tabularnewline \hline
    	\end{tabular}
    	\caption{\label{tbl:kpbofS7FG} Summary of the quantum $k$-producible bounds of $\I_{{\rm S}_n}$ and $\I_{{\rm FG}_n}$, $n=4,\ldots,6$. In contrast with Ref.~\cite{Lin:PRA:2019}, the local bounds of $I_{{\rm FG}_n}$ are normalized to be one here.}
	\end{table}
	
\section{Statistical features of the various visibility distributions}
    
    We summarize in Tables~\ref{vis_MABK}, \ref{tbl:lbvis_state}, and \ref{Tbl:Vis-2-sep}  the statistical properties of the visibility distributions obtained in this work. Included in each table is the maximum, the minimum, the mean, and the standard deviation $\sigma$, as well as the mode of each distribution. To determine the mode (i.e., the most frequently observed value) of a distribution, first we round every visibility value to its second decimal place, and then we search for the mode accordingly. The mode found as such would correspond roughly to the position of the peak of the histogram.
    
    \begin{table}[h!bt]
        \centering
        \begin{tabular}{|c|c|c|c|c|c|c|c|}
            \hline
            $n$ & $\Npt (10^3)$ & Max & Min & Mean & Mode & $\sigma$&Within 1 $\sigma$ \tabularnewline \hline
            3 & 4000 & 1.060 & 0.503 & 0.720 & 0.70 & 0.083 &65.69\% \tabularnewline \hline
            4 & 4000 & 0.880 & 0.359 & 0.553 & 0.55 & 0.066& 66.10\%  \tabularnewline \hline
            5 & 2000 & 0.678 & 0.257 & 0.427 & 0.42 & 0.055 & 66.94\%  \tabularnewline \hline
            6 & 467 & 0.558 & 0.190 & 0.330 & 0.33 & 0.045 & 67.48\%  \tabularnewline \hline
            7 & 450 & 0.445 & 0.136 & 0.254 & 0.25 & 0.037 & 67.94\%  \tabularnewline \hline
            8 & 125 & 0.359 & 0.103 & 0.196 & 0.19 & 0.030 & 68.254\%  \tabularnewline \hline
        \end{tabular}
            \caption{
            Summary of the visibility distributions to $\Lnmk{n}{3}{2}$ obtained by considering the Bell value of all input-lifting of $\Imabk$ to the $(n,3,2)$ Bell scenario and evaluated according to~\cref{Eq:VisIneq}.}
        \label{vis_MABK}
    \end{table}

    \begin{table}[h!bt]    
        \begin{tabular}{|c|c|c|c|c|c|c|c|}
            \hline
             Bell scenario & $\Npt (10^3)$ & Max & Min & Mean & Mode & $\sigma$&Within 1 $\sigma$ \tabularnewline \hline
            Lift (3,2,2) &100&0.823 &0.507&0.696&0.73&0.055&68.6\% \tabularnewline \hline
            (3,3,2) &4000& 0.774&0.503&0.646&0.65&0.03&72.6\%  \tabularnewline \hline
            Lift (4,2,2) & 100 & 0.679 & 0.367 & 0.541 & 0.58 & 0.051 & 65.3\%  \tabularnewline \hline
            (4,3,2) &600&0.620 &0.358&0.512&0.53&0.032&71.4\%  \tabularnewline \hline
            (5,3,2) &4.4 &0.510&0.269&0.406&0.43&0.041&65.7\%  \tabularnewline \hline
        \end{tabular}
            \caption{Summary of the visibility distributions to the various sets of $\Lnmk{n}{3}{2}$ or their approximations. Those listed to the right of a ``Lift $(n,2,2)$" Bell scenario correspond to the case where we consider an approximation to $\Lnmk{n}{3}{2}$ by considering only all input liftings of Bell facets originally defined for $\Lnmk{n}{2}{2}$. See ~\cref{Sec:ProbViolation} and~\cref{Sec:3Paths} for details.
            }
            \label{tbl:lbvis_state}
    \end{table}            

    \begin{table}[h!bt]        	
        \begin{tabular}{|c|c|c|c|c|c|c|c|}
            \hline
             & $\Npt (10^3)$ & Max & Min & Mean & Mode & $\sigma$& Within 1 $\sigma$ \tabularnewline \hline
            Lifting $\S_{3,2}^{(1)}$ &250 & 0.969 &0.711&0.839&0.849&0.031&71.2\% \tabularnewline \hline
            $\S_{3,2}^{(1)}$ &1000 & 0.893&0.704&0.740&0.725&0.018&73.2\%  \tabularnewline \hline
            Lifting $\S_{4,3}^{(1)}$ & 16 & 0.890 & 0.733 & 0.8365 & 0.838 & 0.022 & 68.9\%  \tabularnewline \hline
            $\S_{4,3}^{(1)}$ &10 &0.781 &0.690&0.718&0.701&0.013&63\%  \tabularnewline \hline
        \end{tabular}
            \caption{\label{Tbl:Vis-2-sep} Summary of the visibility distributions to several approximations of the $(n-1)$-producible set for $n=3$ and 4.}
    \end{table}

\clearpage
\end{document}